\documentclass[prd,superscriptaddress,amsfonts,amssymb,onecolumn,amsmath,showpacs]{revtex4-2}
\usepackage{bm}
\usepackage{amsfonts}
\usepackage{tensor}

\usepackage{latexsym}
\usepackage[utf8]{inputenc}
\usepackage{graphicx}
\usepackage{amsmath}
\usepackage{palatino}
\usepackage{mathpazo}
\usepackage[british]{babel}
\usepackage{hhline}
\usepackage{multirow}
\usepackage{textcomp}
\usepackage{float}
\usepackage{booktabs}
\usepackage{dcolumn}
\usepackage{lipsum} 
\usepackage{mdframed}
\usepackage[]{mdframed}
\usepackage{hhline}
\usepackage{multirow}
\usepackage{ragged2e}
\usepackage{hyperref}
\hypersetup{colorlinks,citecolor=blue}
\hypersetup{colorlinks=true,linkcolor=red,filecolor=magenta,    urlcolor=cyan}
\usepackage{amsmath}
\usepackage{xcolor}
\usepackage{orcidlink}
\usepackage{epsfig}
\usepackage{caption}
\usepackage{subcaption}
\usepackage{commath}
\def\jnl@style{\it}
\def\aaref@jnl#1{{\jnl@style#1}}

\def\aaref@jnl#1{{\jnl@style#1}}

\def\aj{\aaref@jnl{AJ}}                   % Astronomical Journal
\def\apj{\aaref@jnl{ApJ}}                 % Astrophysical Journal
\def\apjl{\aaref@jnl{ApJ}}                % Astrophysical Journal, Letters
\def\apjs{\aaref@jnl{ApJS}}               % Astrophysical Journal, Supplement
\def\apss{\aaref@jnl{Ap\&SS}}             % Astrophysics and Space Science
\def\aap{\aaref@jnl{A\&A}}                % Astronomy and Astrophysics
\def\aapr{\aaref@jnl{A\&A~Rev.}}          % Astronomy and Astrophysics Reviews
\def\aaps{\aaref@jnl{A\&AS}}              % Astronomy and Astrophysics, Supplement
\def\mnras{\aaref@jnl{Mon.~Not.~Roy.~Astron.~Soc.}}             % Monthly Notices of the RAS
\def\prd{\aaref@jnl{Phys.~Rev.~D}}        % Physical Review D
\def\prc{\aaref@jnl{Phys.~Rev.~C}}  % Physical Review C
\def\prl{\aaref@jnl{Phys.~Rev.~Lett.}}    % Physical Review Letters
\def\qjras{\aaref@jnl{QJRAS}}             % Quarterly Journal of the RAS
\def\skytel{\aaref@jnl{S\&T}}             % Sky and Telescope
\def\ssr{\aaref@jnl{Space~Sci.~Rev.}}     % Space Science Reviews
\def\zap{\aaref@jnl{ZAp}}                 % Zeitschrift fuer Astrophysik
\def\nat{\aaref@jnl{Nature}}              % Nature
\def\aplett{\aaref@jnl{Astrophys.~Lett.}} % Astrophysics Letters
\def\apspr{\aaref@jnl{Astrophys.~Space~Phys.~Res.}} % Astrophysics Space Physics Research
\def\physrep{\aaref@jnl{Phys.~Rep.}}      % Physics Reports
\def\physscr{\aaref@jnl{Phys.~Scr}}       % Physica Scripta
\def\commat{\aaref@jnl{Comm.~Math.~Phys.}}              % Communications in Mathematical Physics
\def\science{\aaref@jnl{Science}}               % Science
\def\cqg{\aaref@jnl{Classical Quant.~Grav.}}            % Classical and Quantum Gravity
\def\jpcs{\aaref@jnl{JPCS}}                                     % Journal of Physics Conference Series
\def\ijmpd{\aaref@jnl{Int.~J.~Mod.~Phys.~D}}                    % International Journal of Modern Physics D
\def\grg{\aaref@jnl{Gen.~Relat.~Gravit.}}               % General Relativity and Gravitation
\def\rpp{\aaref@jnl{Rep.~Prog.~Phys.}}          % Reports on Progress in Physics
\def\npa{\aaref@jnl{Nucl.~Phys.~A}}        % Nuclear Physics A
\def\lrr{\aaref@jnl{Living Rev.~Rel.}}                   % Living reviews in relativity
\def\jcap{\aaref@jnl{J.~Cosmology Astropart.~Phys.}}    % Journal of cosmology and astroparticle physics
\def\rmp{\aaref@jnl{Rev.~Mod.~Phys.}}   %Reviews of modern physics
\def\epjc{\aaref@jnl{Eur.~Phys.~J.~C}} 
\def\plb{\aaref@jnl{~Phy.~Lett.~B}} 
\def\mpla{\aaref@jnl{Mod.~Phy.~Lett.~A}} 
\def\arxiv{\aaref@jnl{arxiv.org}}

\allowdisplaybreaks[1]
\begin{document}

\color{black}       %% For one column

\title{Bouncing Cosmological Models and Energy Conditions in $f(Q, L_m)$ gravity}

\author{S. A. Kadam\orcidlink{0000-0002-2799-7870}}
\email{siddheshwar.kadam@dypiu.ac.in;
\\k.siddheshwar47@gmail.com}
\affiliation{Centre for Interdisciplinary Studies and Research, D Y Patil International University, Akurdi, Pune-411044, Maharashtra, India}

\author{V. A. Kshirsagar\orcidlink{0009-0003-1256-2246}}
\email{kvitthal99@gmail.com}
\affiliation{JSPM's Bhivarabai Sawant Institute of Technology and Research, Wagholi, Pune-412207, Maharashtra, India}
\author{Santosh Kumar Yadav\orcidlink{0009-0009-2581-387X}}
\email{sky91bbaulko@gmail.com}
\affiliation{Department of Mathematics, SR University, Warangal-506371,
Telangana, India.}
\begin{abstract}
% In this work, the bouncing solutions in modified $f(Q, L_m)$
% gravity have been analysed. The four well-known bouncing models, the symmetric bounce, super bounce, oscillatory bounce, and matter bounce, widely studied in the modified gravity formalism, are analysed. The nature of the Hubble parameter, evolution of the scale factor, and equation of state parameters are investigated. The behaviour of scale factor and the Hubble parameter adress the bouncing scenario successfully. The EoS parametr at bouncing epoch lies into the phantom region supporting the bounce. To investigate the successful description of the bouncing scenario, the energy conditions have been tested at each bouncing model. The violation of null energy condition experiences at the bouncing epoch, which succesfully adress the bouncing behaviour of the model.
\textbf{Abstract:} This study explores the bouncing solutions within the framework of modified \( f(Q, L_m) \) gravity. We examine four prominent bouncing models: the symmetric bounce, super bounce, oscillatory bounce, and matter bounce, each of which has been extensively analyzed in the context of modified gravity theories. Our investigation focuses on the behavior of the Hubble parameter, the evolution of the scale factor, and the equation of state (EoS) parameters. Notably, the dynamics of the scale factor and Hubble parameter effectively support the bouncing scenario. During the bouncing epoch, the EoS parameters fall within the phantom region, reinforcing the viability of the bounce. To further validate the bouncing scenario, we assess the energy conditions associated with each model. Our findings reveal a violation of the null energy condition at the bouncing epoch, which successfully characterizes the model’s bouncing behavior.
\end{abstract}

\maketitle

\textbf{Keywords}: $f(Q, L_m)$  gravity, bouncing cosmology, energy conditions

\section{Introduction}\label{Introduction}

The theoretical discussions and observational evidence suggest that the universe underwent an early phase of expansion known as inflation \cite{Guth1981, Weinberg2008,Carroll1992}. The key point of the inflationary model is that it is particularly effective in addressing the flatness, horizon, and monopole problems \cite{Linde:1981mu}. It also provides a coherent framework for the generation of primordial fluctuations and primordial gravitational waves \cite{Peter_2002,Farrugia:2018gyz}. As a result, current observational studies \cite{Riess:1998cb, Perlmutter:1998np} are focused on uncovering the evolution of the Universe across all epochs. The theoretical investigations aim to incorporate the remarkable new findings into a coherent theoretical framework, preferably within a unified theory \cite{Di_Valentino_2025}. An alternative to the typical inflationary explanation for the early acceleration is offered by bouncing cosmologies \cite{Lohakare_2022,agrawal2022bouncing}, which eliminate the occurrence of an initial singularity. Bouncing cosmology presents an intriguing approach to tackling the problems related to initial singularities \cite{Gul2024}. This idea seeks to address the issues posed by the big bang singularity, which is a notable challenge in cosmology \cite{Singh:2022gln,Bamba2014}. The core concept of bouncing cosmology is to suggest a model in which the universe does not originate from a singular point (as proposed by the Big Bang) but rather experiences a contraction followed by a bounce, leading to its current expansion \cite{Ashtekar_2006}. This framework helps to circumvent the theoretical difficulties and infinite values associated with singularities, providing a more seamless and precise explanation of the cosmic origin and its dynamic characteristics \cite{Odintsov2016}. 

In bouncing cosmology, the Universe transitions from an initial phase of contraction to a later phase of expansion at the point of the bounce. The Hubble parameter $H(t)$ shifts from a negative region to a positive region, that is, $(H(t) < 0$ to $H(t) > 0)$. At the bounce point, the Hubble parameter vanishes \cite{Cai2007}. In this study, a spatially flat Friedmann-Lemaître-Robertson-Walker (FLRW) universe has been considered. In this setup, it is well established that the Null Energy Condition for matter must be violated to successfully address the bouncing phenomenon. This is because even though the Hubble parameter is zero at the bounce point, its time derivative remains positive. Consequently, the EoS parameter, $\omega$, of the gravity theory is less than $-1$ \cite{Brown_2008}. The bouncing phenomena have been analysed widely in the different gravity representations, like modifications to GR \cite{Lohakare_2022,Bamba_2014,Singh2023}, teleparallel \cite{Kadam_bouncing_2026,Mishra_2024_bouncing,caruana2020,dela_2018}, and symmetric teleparallel gravity formalisms \cite{Gul2024}.

In the present formalism, the symmetric teleparallel gravity, in which nonmetricity alters the torsion and the metric affine connection, has been explored \cite{BeltranJimenez_2017}. The cosmic bounce has been explored in $f(Q)$ gravity, which is one of the immediate modifications to symmetric teleparallel gravity, as explored in \cite{Bajardi2020}, which tackles several concerns of the early universe. The bouncing solutions and the investigation of the behavior of the scalar perturbation are performed to check the stability in $f(Q)$ gravity \cite{Koussour2024,Bajardi2020}. The non-singular matter bounce scenario is investigated through the cosmographic tests in the further modification to $f(Q)$ gravity by adding the trace of the energy momentum tensor in $f(Q, T)$, the formalism \cite{AGRAWAL2021100863}. The reconstruction of the Hubble parameter is also performed to study the matter bounce scenario in $f(Q, T)$ gravity in Ref. \cite{Sharif2024}. The comparative studies along with the analytical solutions are investigated in the higher-order symmetric teleparallel gravity formalism with the inclusion of a boundary term in $f(Q, C)$ gravity, which can be reviewed in \cite{Ghosh2026,Samaddar2025,Samaddar:2026mda}. Motivated by the comprehensive analysis of the different bouncing solutions that have been previously performed in modified teleparallel \cite{caruana2020} and symmetric teleparallel gravity theories \cite{Samaddar:2026mda}, including boundary terms. This analysis presents a comprehensive analysis of the four important bouncing cosmological models in the framework of $f(Q, L_m)$ gravity \cite{Hazarika:2024alm}. The comprehensive analysis of the Bianchi I Universe and the influence of bulk viscosity on the viability of cosmological bounce solutions in the $f(Q, L_m)$ is studied in \cite{Sharif2025b}. Further, the role of $f(Q, L_m)$ gravity formalism in the presence of bouncing parameterisation and the analysis of different cosmographic parameters have been studied \cite{Sharif2025}.

This paper is organised as follows: Section \ref{formalism} provides details on the background and fundamental formalisms required to frame the $f(Q, L_m)$ gravity formalism. In Section \ref{bouncing_models}, the four well-known bouncing solutions, symmetric bounce, superbounce, oscillatory bounce, and matter bounce, have been presented. In Section \ref{energy_conditions}, the energy conditions have been investigated to check the viability of the occurrence of the bounce. Finally, in Section \ref{conclsion}, the crucial findings have been discussed.

\section{FORMALISM OF $f(Q, L_m)$ GRAVITY THEORY}\label{formalism}
In the $f(Q, L_m)$ modified gravity formalism, the gravitational action is constructed by covering the Lagrangian density as a function of the scalar non-metricity $Q$ and the Lagrangian of matter $L_m$. The action is written as \cite{Hazarika:2024alm},
\begin{equation}\label{frl1}
    S=\int f(Q,L_m) \sqrt{-g} d^4x \,,
\end{equation}
where, $g=\det(g_{\mu\nu})$ is the determinant of the metric tensor and $\sqrt{-g}$ confirms the general covariance of the action. The arbitrary function $f(Q, L_m)$ denotes the coupling between the geometric sector, defined by the scalar of non-metricity $Q$, and the matter sector, described by the Lagrangian of matter $L_m$.

The non-metricity scalar $Q$ is given by \cite{BeltranJimenez_2017},
\begin{equation}
     Q \equiv -g^{\mu\nu} \left(L^{\beta}{}_{\alpha\mu} \, L^{\alpha}{}_{\nu\beta} - L^{\beta}{}_{\alpha\beta} \, L^{\alpha}{}_{\mu\nu} \right) \,, 
\end{equation}
where, $L^{\beta}{}_{\alpha\gamma}$ is the disformation tensor, defined as
\begin{equation}
    L^{\beta}{}_{\alpha\gamma} = \frac{1}{2} \, g^{\beta\eta} \left( Q_{\gamma\alpha\eta} + Q_{\alpha\eta\gamma} - Q_{\eta\alpha\gamma} \right)\,.
\end{equation}
The non-metricity tensor, which measures the failure of the metric to remain covariantly conserved, is defined as
\begin{equation}
    Q_{\gamma\mu\nu} \equiv -\nabla_{\gamma} g_{\mu\nu} = -\partial_{\gamma} g_{\mu\nu} + g_{\nu\sigma}\,\tilde{\Gamma}^{\sigma}{}_{\mu\gamma} + g_{\sigma\mu}\,\tilde{\Gamma}^{\sigma}{}_{\nu\gamma}\,,
\end{equation}
where, $\tilde{\Gamma}^{\gamma}{}_{\mu\nu}$  denotes the affine connection in symmetric teleparallel geometry. 

The traces of the non-metricity tensor are expressed as,
\begin{equation}
    Q_{\beta} = g^{\mu\nu} Q_{\beta\mu\nu}, \qquad \tilde{Q}_{\beta} = g^{\mu\nu} Q_{\mu\beta\nu}\,.
\end{equation}
To simplify the derivation of the field equations, we familiarize the superpotential tensor (also known as the non-
metricity conjugate) \cite{DAgostino:2018ngy}, defined as
\begin{equation}
    P^{\beta}{}_{\mu\nu} \equiv \frac{1}{4} \left[- Q^{\beta}{}_{\mu\nu}+ 2 Q_{(\mu}{}^{\beta}{}_{\nu)}+ Q^{\beta} g_{\mu\nu}- \tilde{Q}^{\beta} g_{\mu\nu}- \delta^{\beta}_{(\mu} Q_{\nu)}\right]\,,
\end{equation}
which can be equivalently  written as
\begin{equation}
    P^{\beta}{}_{\mu\nu}= -\frac{1}{2} L^{\beta}{}_{\mu\nu}+ \frac{1}{4} \left( Q^{\beta} - \tilde{Q}^{\beta} \right) g_{\mu\nu}- \frac{1}{4} \delta^{\beta}_{(\mu} Q_{\nu)}\,.
\end{equation}
This tensor plays a role comparable to the contortion tensor in teleparallel gravity. The non-metricity scalar can be communicated in terms of this conjugate tensor as \cite{BeltranJimenez_2017}
\begin{equation}
    Q = - Q_{\beta\mu\nu} P^{\beta\mu\nu}= -\frac{1}{4} \left(-  Q^{\beta\nu\rho} Q_{\beta\nu\rho} + 2 Q^{\beta\nu\rho} Q_{\rho\beta\nu} - 2 Q^{\rho} \tilde{Q}_{\rho}+ Q^{\rho} Q_{\rho} \right)\,.
\end{equation}
By varying the action (\ref{frl1}) with respect to the metric tensor, the gravitational field equations are achieved as
\begin{equation} \label{eq.9}
    \frac{2}{\sqrt{-g}} \nabla_{\alpha} \left( f_Q \sqrt{-g}\, P^{\alpha}{}_{\mu\nu} \right)+ f_Q \left( P_{\mu\alpha\beta} Q_{\nu}{}^{\alpha\beta}- 2 Q^{\alpha\beta}{}_{\mu} P_{\alpha\beta\nu} \right)+ \frac{1}{2} f\, g_{\mu\nu}= \frac{1}{2} f_{L_m} \left( g_{\mu\nu} L_m - T_{\mu\nu} \right) \tag{9}\,,
\end{equation}
where,
\[ f_Q = \frac{\partial f(Q, L_m)}{\partial Q}, \qquad
f_{L_m} = \frac{\partial f(Q, L_m)}{\partial L_m}.
\]
For this particular choice of function $f(Q, L_m) = f(Q) + 2L_m$, the above field equations take the form of field equations for $f(Q)$ gravity \cite{BeltranJimenez_2017}. The energy--momentum tensor of matter is defined as,
\begin{equation}
  T_{\mu\nu}= -\frac{2}{\sqrt{-g}} \frac{\delta \left( \sqrt{-g}\, L_m \right)}{\delta g^{\mu\nu}}= g_{\mu\nu} L_m - 2 \frac{\partial L_m}{\partial g^{\mu\nu}}\,.
\end{equation}
Again, varying the action with respect to the affine connection produces an additional field equation,
\begin{equation}
    \nabla_{\mu} \nabla_{\nu} \left( 4 \sqrt{-g}\, f_Q \, P^{\mu\nu}{}_{\alpha} + H_{\alpha}{}^{\mu\nu} \right) = 0, \tag{10}
\end{equation}
where,  $H_{\alpha}{}^{\mu\nu}$ denotes the hypermomentum density, defined as
\begin{equation}
    H^{\mu\nu}{}_{\alpha} = \sqrt{-g}\, f_{L_m} \, \frac{\delta L_m}{\delta Y^{\alpha}{}_{\mu\nu}}\,.
\end{equation}
An important characteristic of the $ f(Q, L_m) $ gravity theory is the non-conservation of the energy-momentum tensor. By taking the covariant derivative of the field equations (\ref{eq.9}), 
\begin{equation}
    \nabla_{\mu} T^{\mu}{}_{\nu}= \frac{1}{f_{L_m}} \left[\frac{2}{\sqrt{-g}} \nabla_{\alpha} \nabla_{\mu} H_{\nu}{}^{\alpha\mu}+ \nabla_{\mu} A^{\mu}{}_{\nu}- \nabla_{\mu} \left( \frac{1}{\sqrt{-g}} \nabla_{\alpha} H_{\nu}{}^{\alpha\mu} \right)\right]\equiv B_{\nu} \neq 0\,, 
\end{equation}

which evidently specifies the non-conservation of the energy-momentum tensor.
The non-zero term $B_{\nu}$ indicates the flow of energy and momentum between the matter sector and the geometric sector, as a result of the non-minimal coupling.

\section{FIELD EQUATIONS OF MODIFIED $f(Q,Lm)$ GRAVITY}
In order to study cosmological dynamics within the \( f(Q, L_m) \) gravity theory, we take a spatially flat FLRW spacetime. This theory is based on the cosmological principle of homogeneity and isotropy, which states that our Universe is homogeneous on a large scale and has no preferred directions \cite{Ryden_2016}. Due to such symmetries, a FLRW spacetime gives us a suitable description of an expanding Universe. 

The spacetime geometry of a FLRW Universe can be described by a line element given by
\begin{equation} \label{eq.12}
    ds^{2} = -dt^{2} + a^{2}(t)\left(dx^{2} + dy^{2} + dz^{2}\right)\,,
\end{equation}

here, $a(t)$ represents a scale factor, which is a time-dependent function.
In this model of the universe, the scalar of non-metricity takes a mainly simple form given by
$Q = 6H^{2}$,
where the Hubble parameter is given by
$H = \frac{\dot{a}}{a}$,
which deals with the rate at which the universe expands.

To describe the matter content of the universe, we consider it to be a perfect fluid and the corresponding energy--momentum tensor is given by
\begin{equation} \label{eq.13}
    T_{\mu\nu} = (\rho + p)\, u_{\mu} u_{\nu} + p\, g_{\mu\nu}\,,
\end{equation} 
where $\rho$ and $p$ denote the energy density and pressure of the fluid, respectively, and $u_{\mu}$ denotes the four-velocity of the fluid, satisfying the condition $u_{\mu} u^{\mu} = -1$.

With the FLRW metric given by Eq. (\ref{eq.12}) and the energy-momentum tensor for a perfect fluid given by Eq. (\ref{eq.13}), and by applying the variational principle to the modified gravity action \cite{Hazarika:2024alm}, the modified Friedmann equations for $f(Q, L_m)$ gravity can be established in the following forms:
\begin{align}
3H^{2} &= \frac{1}{4 f_Q} \left[ f - f_{L_m} (\rho + L_m) \right]\,, \label{eq.14} \\
\dot{H} + 3H^{2} + \frac{\dot{f}_Q}{f_Q} H
&= \frac{1}{4 f_Q} \left[ f + f_{L_m} (p - L_m) \right]\,. \label{eq.15}
\end{align}
The model we explored in this study is $f(Q, L_m)=\beta +\alpha  L_m Q^{\mu }-\frac{Q}{2}$ \cite{Hazarika:2024alm}, which has been successfully tested to validate the energy conditions \cite{Myrzakulov_2025_EC} and some other important cosmological aspects like baryogenesis and the observational data analysis \cite{Amit_NPB,Myrzakulov_2025}. For the above particular choice of a functional form of $f(Q,L_m)$, one finds that
\begin{align}
f_Q &= \frac{-1}{2}+\alpha\mu L_{m} Q^{\mu-1}\,, \qquad f_{L_{m}} = \alpha Q^{\mu}\,.
\end{align}
Taking into account a Lagrangian density of matter in the form $L_{m}=\rho$ \cite{Harko:2015pma} 
the modified Friedmann equations leading the cosmological evolution of a Universe occupied with bulk viscous matter are instantly obtained from Eqs. (\ref{eq.14}) and (\ref{eq.15}) in the following form:
\begin{align}
p &= -\frac{2^{-\mu } 3^{-\mu -1} \left(H^2\right)^{-\mu -1} \left(9 H^4+3 \beta  H^2+6 \dot{H} H^2 \mu +6 \dot{H} H^2+2 \beta  \dot{H} \mu \right)}{\alpha  (2 \mu +1)}\,, \\
\rho &= \frac{6^{-\mu } \left(\beta +3 H^2\right) \left(H^2\right)^{-\mu }}{\alpha  (2 \mu +1)}\,, \\
\omega &= \frac{2^{-\mu } 3^{-\mu -1} \left(H^2\right)^{-\mu -1} \left(9 H^4+3 \beta  H^2+6 \dot{H} H^2 \mu +6 \dot{H} H^2+2 \beta  \dot{H} \mu \right)}{6^{-\mu } \left(\beta +3 H^2\right) \left(H^2\right)^{-\mu }}\,.
\end{align}
\section{Bouncing Models}\label{bouncing_models}
\subsection{Model I: Symmetric Bounce Scale Factor}

\begin{figure}[H]
\centering

\begin{subfigure}{0.32\textwidth}
\centering
\includegraphics[width=\linewidth]{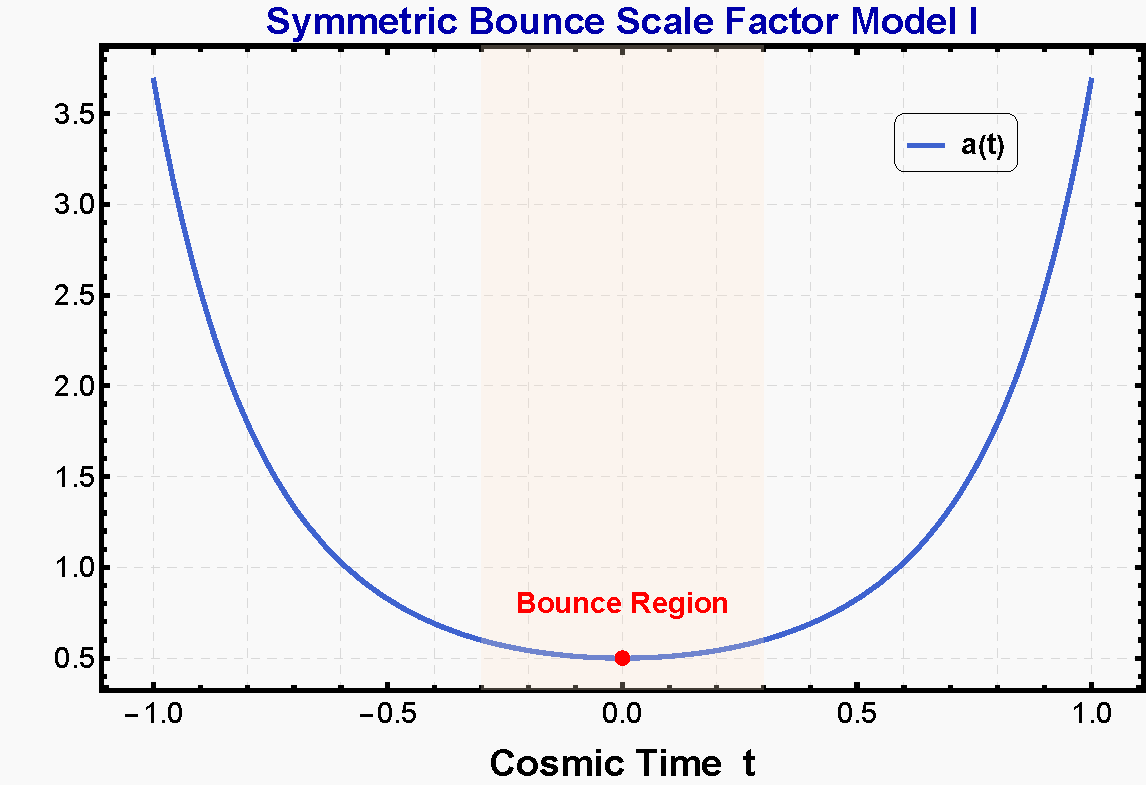}
\caption{Symmetric bounce scale factor}
\end{subfigure}
\hfill
\begin{subfigure}{0.32\textwidth}
\centering
\includegraphics[width=\linewidth]{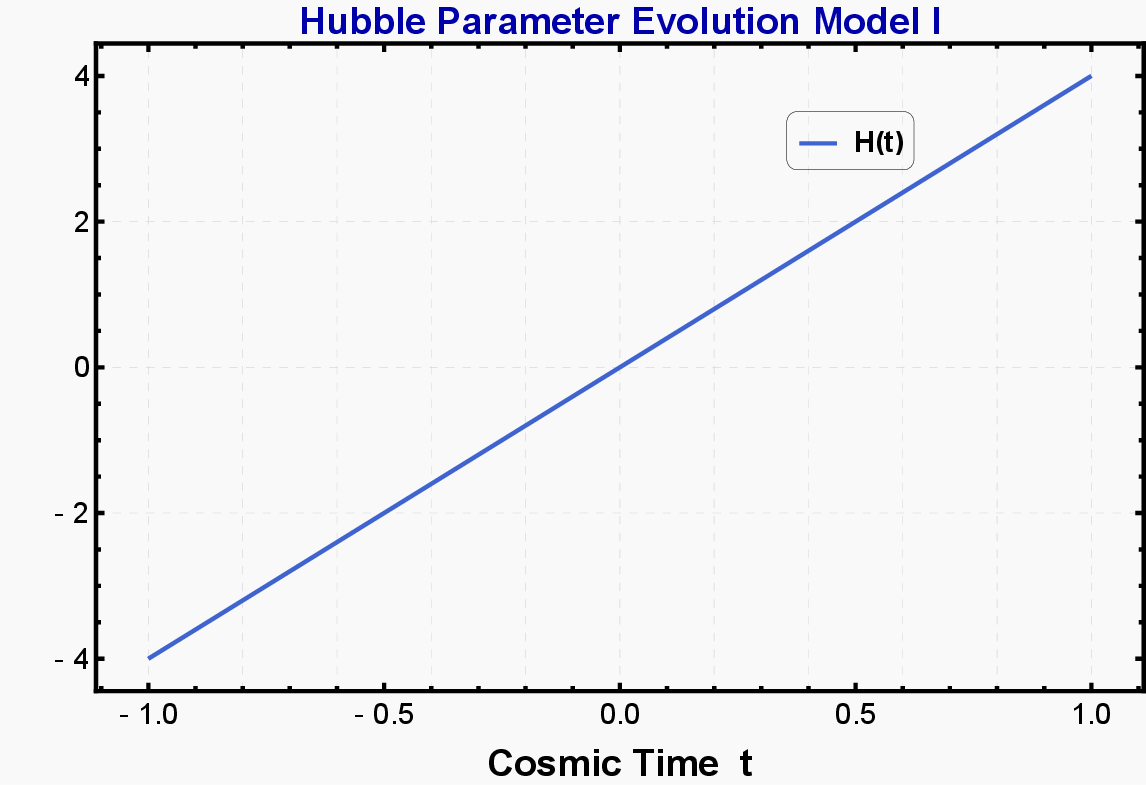}
\caption{Hubble parameter}
\end{subfigure}
\hfill
\begin{subfigure}{0.32\textwidth}
\centering
\includegraphics[width=\linewidth]{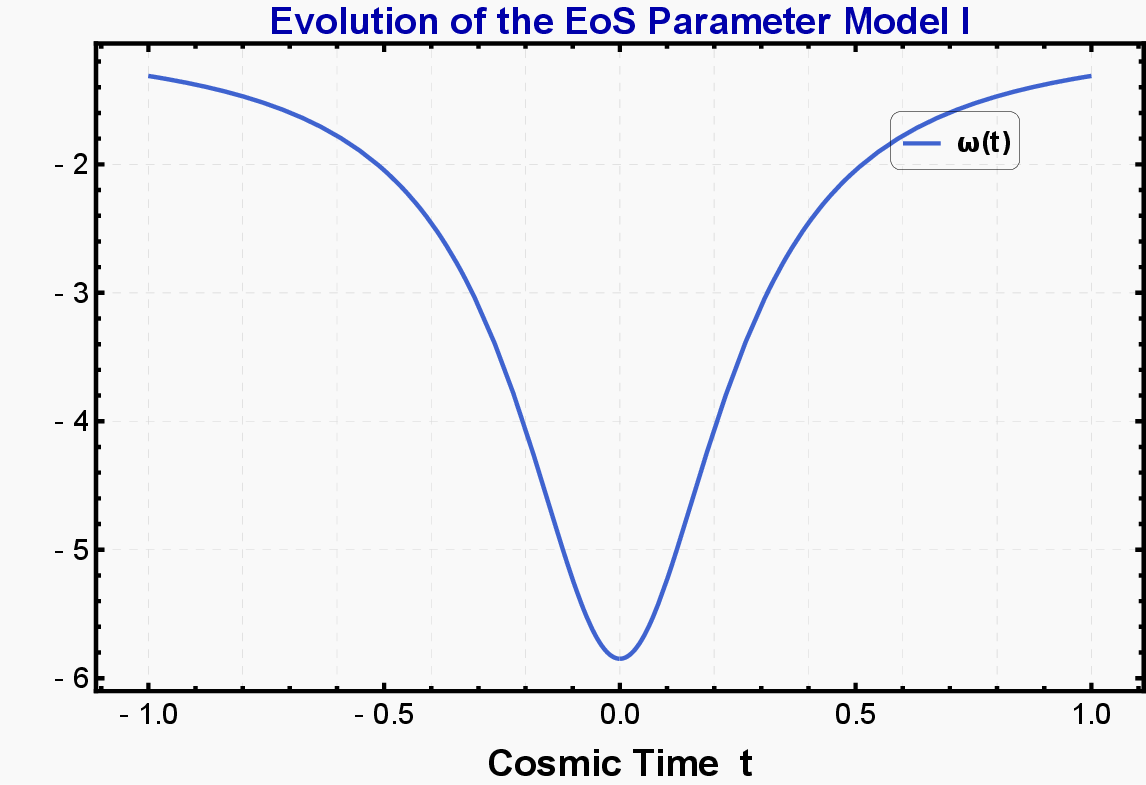}
\caption{EoS parameter}
\end{subfigure}
\caption{Evolution of the scale factor, Hubble parameter, and EoS parameter for Model I with the symmetric bounce scale factor.}
\label{fig:model1}
\end{figure}

Our first model is the symmetric bounce, which was first investigated in \cite{Cai_2012}. The form of the scale factor in this case is $a(t)=A \exp \left(\frac{B t^2}{T^2}\right)$, where $A, B, T$ the arbitrary constants with the condition $A, B>0$. The constant $T$ can be treated as an arbitrary time. The Hubble parameter in this case will take the form $\frac{2 B t}{T^2}$, with the nonmetricity scalar $Q=\frac{24 B^2 t^2}{T^4}$. Clearly, the bounce occurs at $t = 0$, which is preceded by a contracting phase $(t < 0)$ and followed by an expanding phase $(t > 0)$, as shown in Fig. \ref{fig:model1}. The EoS parameter at the bounce lies in the phantom region, which successfully addresses the bouncing behaviour in $f(Q, L_m)$ gravity theory. In this context, it's evident that the evolution rate of the Universe $a(t)$ Fig. \ref{fig:model1} (a), the Hubble parameter $H(t)$ \ref{fig:model1} (b), and the EoS parameter \ref{fig:model1} (c) demonstrate a symmetric nature. The bouncing region in the scale factor diagram is highlighted with the light orange strip.

\subsection{Model II: Superbounce Scale Factor}

\begin{figure}[H]
\centering

\begin{subfigure}{0.32\textwidth}
\centering
\includegraphics[width=\linewidth]{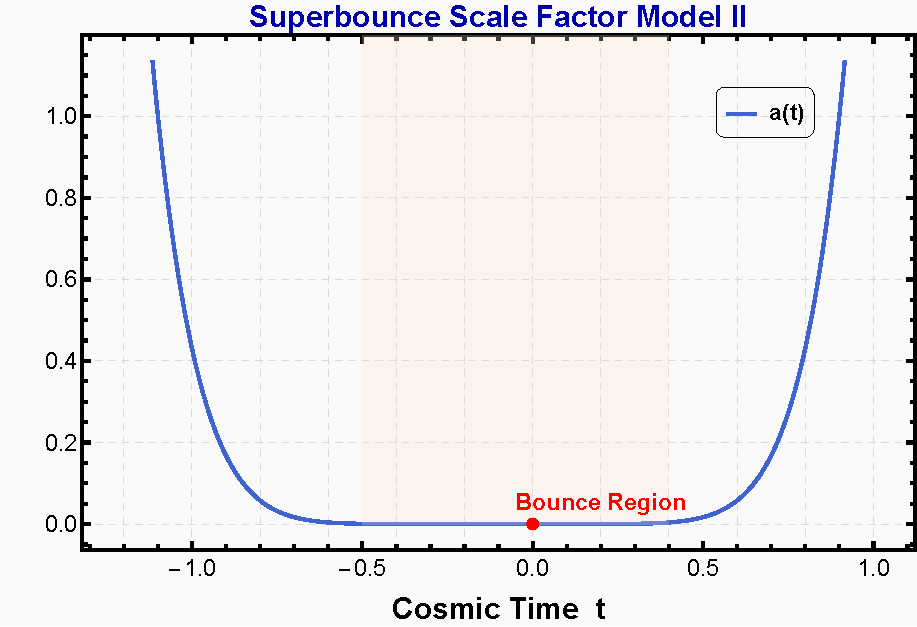}
\caption{Superbounce scale factor}
\end{subfigure}
\hfill
\begin{subfigure}{0.32\textwidth}
\centering
\includegraphics[width=\linewidth]{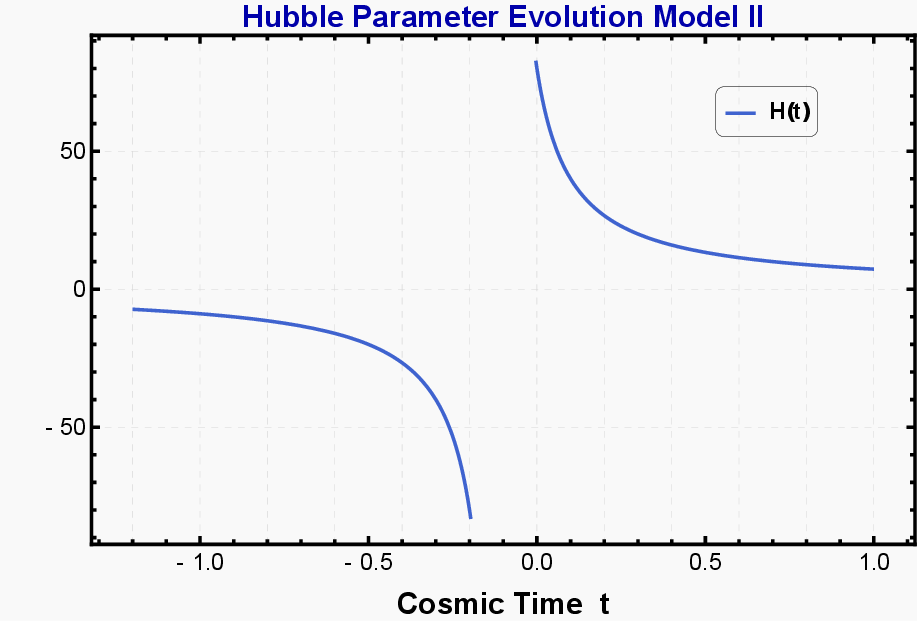}
\caption{Hubble parameter}
\end{subfigure}
\hfill
\begin{subfigure}{0.32\textwidth}
\centering
\includegraphics[width=\linewidth]{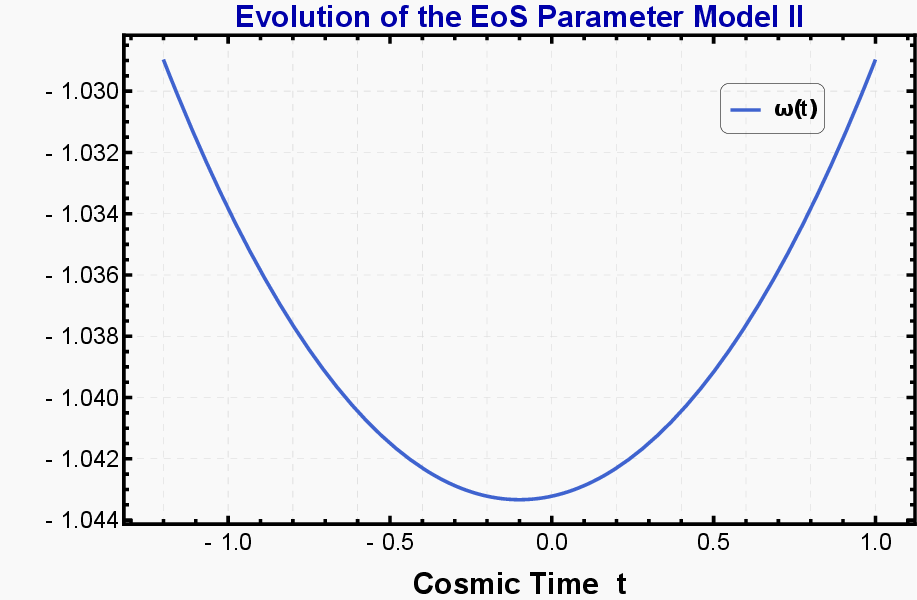}
\caption{EoS parameter}
\end{subfigure}

\caption{Evolution of the scale factor, Hubble parameter, and EoS parameter for Model II with the superbounce scale factor.}
\label{fig:model2}
\end{figure}

The superbounce scale factor will take the form $a(t)=\frac{(t_s-t)}{t_0}^{\frac{2}{c^2}}$, where $c$ is a constant greater than $\sqrt{6}$, $t_s$ represents the time at which the bounce takes place, and $t_0$ is a positive arbitrary time that ensures when $t$ equals $t_s + t_0$, the scale factor takes on a value of one. Here $H=-\frac{2}{c^2 (t_s-t)}$ and $Q=\frac{24}{c^4 (t_s-t)^2}$. This form was originally discussed in \cite{Koehn_2014}. This form is utilized to build a universe that experiences collapse and rebirth through a Big Bang without the presence of a singularity \cite{Oikonomou2015}. The behaviour of the scale factor, the Hubble parameter, which shows a singularity at the bouncing point, ensures the formation of the bouncing nature \ref{fig:model2} (a, b). The Hubble paramter  The EoS parameter in this case, is lying in the phantom region at the bounce epoch.

\subsection{Model III: Oscillatory Bounce Scale Factor}
\begin{figure}[H]
\centering

\begin{subfigure}{0.32\textwidth}
\centering
\includegraphics[width=\linewidth]{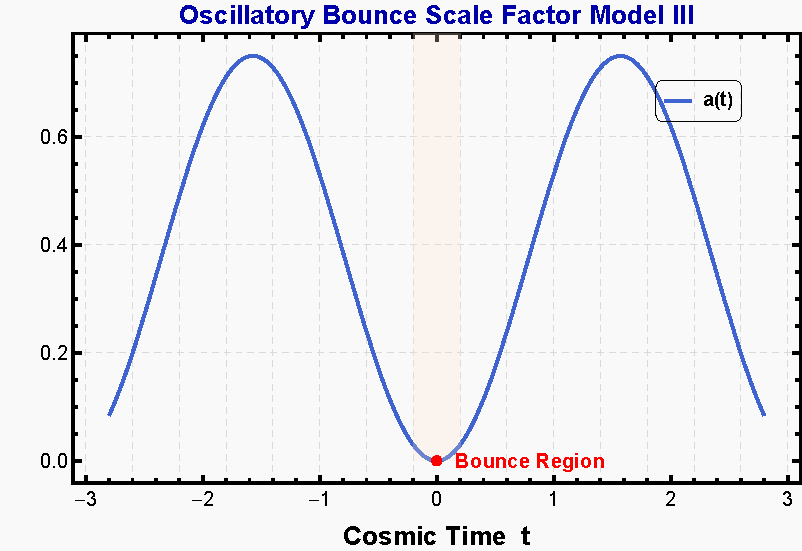}
\caption{Oscillatory bounce scale factor}
\end{subfigure}
\hfill
\begin{subfigure}{0.32\textwidth}
\centering
\includegraphics[width=\linewidth]{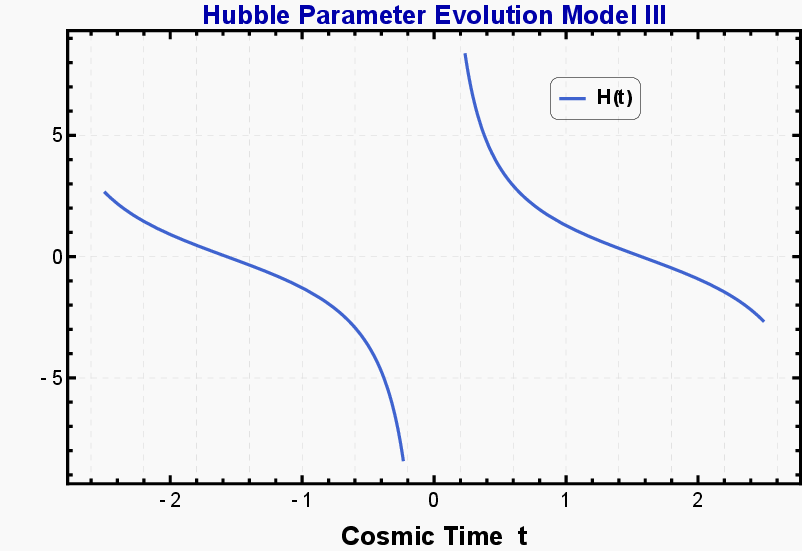}
\caption{Hubble parameter}
\end{subfigure}
\hfill
\begin{subfigure}{0.32\textwidth}
\centering
\includegraphics[width=\linewidth]{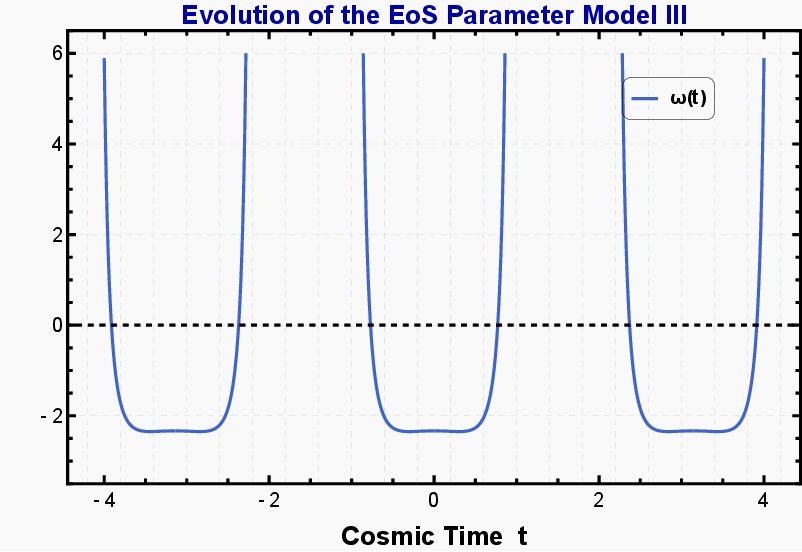}
\caption{EoS parameter}
\end{subfigure}

\caption{Evolution of the scale factor, Hubble parameter, and EoS parameter for Model III with the oscillatory bounce scale factor.}
\label{fig:model3}
\end{figure}

The next form of the scale factor we are taking into consideration is the oscillatory bounce, which takes the form $a(t)=B \sin ^2\left(\frac{\beta  t}{T}\right)$ \cite{NOVELLO_2008}. Here $B, \beta $ are the constants, and the parameter $T$ is the reference time and, for convenience, can be chosen to be $T>0$. The Hubble parameter will take the form, $H(t)=\frac{2 \beta  \cot \left(\frac{\beta t}{T}\right)}{T}$, and the nonmetricity scalar can be read as $Q=\frac{24 \beta ^2 \cot ^2\left(\frac{\beta  t}{T}\right)}{T^2}$. The scale factor refers to Fig. \ref{fig:model3} (a); the bouncing region is highlighted with a light orange color. This behavior allows for the experience of a cyclic Universe. The Hubble parameter experiences the singularity at the bouncing point and can be visualised from Fig. \ref{fig:model3} (b). The EoS parameters \ref{fig:model3} (c) undergo a phase of contraction followed by a phase of expansion, forming a continuous sequence. The EoS lies in the phantom region at the time of the bounce epoch and shows oscillations throughout the evolution.

\subsection{Model IV: Matter Bounce Scale Factor}

\begin{figure}[H]
\centering

\begin{subfigure}{0.32\textwidth}
\centering
\includegraphics[width=\linewidth]{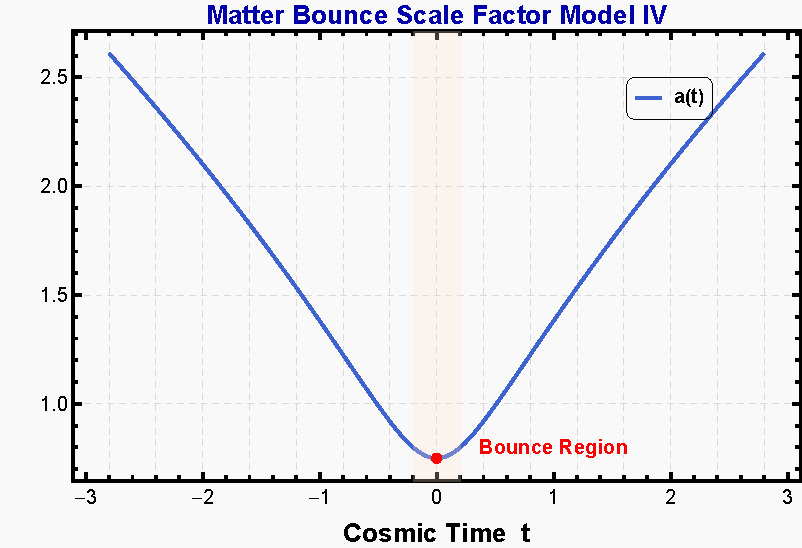}
\caption{Matter bounce scale factor}
\end{subfigure}
\hfill
\begin{subfigure}{0.32\textwidth}
\centering
\includegraphics[width=\linewidth]{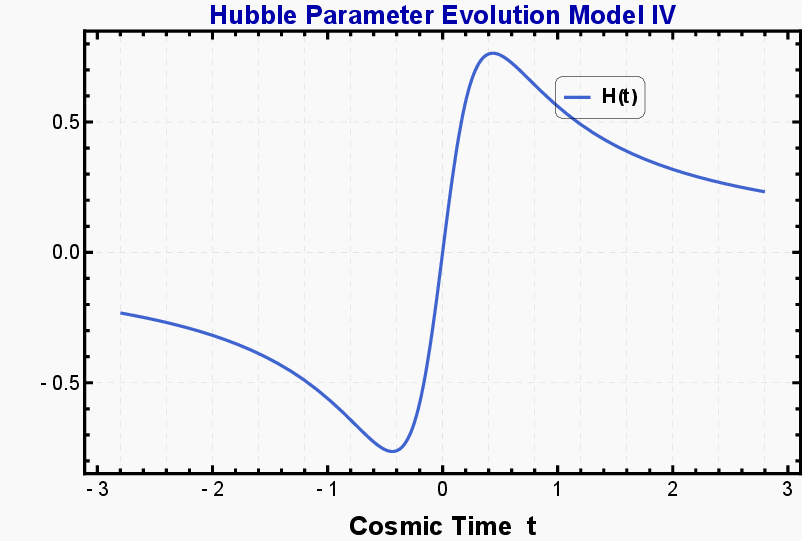}
\caption{Hubble parameter}
\end{subfigure}
\hfill
\begin{subfigure}{0.32\textwidth}
\centering
\includegraphics[width=\linewidth]{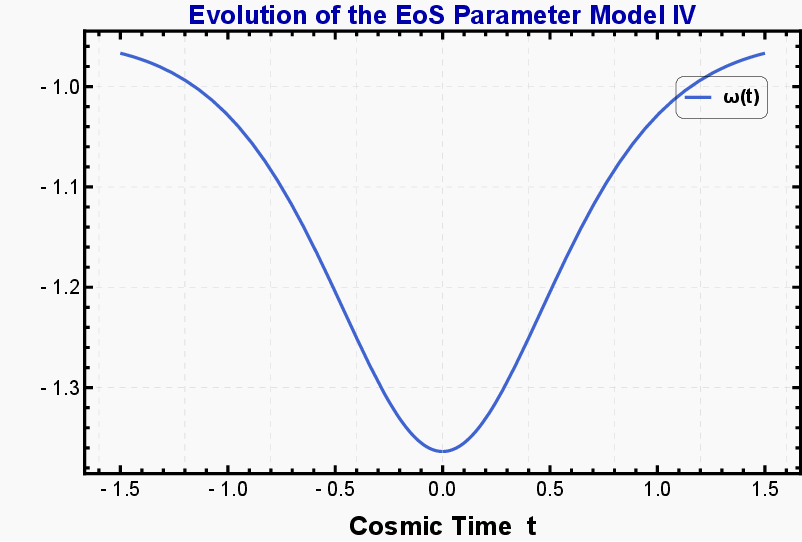}
\caption{EoS parameter}
\end{subfigure}

\caption{Evolution of the scale factor, Hubble parameter, and EoS parameter for Model IV with the matter bounce scale factor.}
\label{fig:model4}
\end{figure}

The next model originates from loop quantum cosmology (LQC) and leads to what is known as matter bounce cosmology \cite{Singh_2006}. This specific form of bouncing cosmology has been explored in the universe's early stages and has demonstrated the potential to create a scale-invariant (or nearly scale-invariant) power spectrum based on the matter fluid being examined \cite{Edward_Wilson2013}. The form of the scale factor in this case is $a(t)=D \sqrt[3]{\frac{3 \tau  t^2}{2}+1}$, where $D>0, 0<\tau\le 1$ are the arbitrary constants. The Hubble parameter and the non-metricity scalar can be presented as respectively $H(t)=\frac{t \tau }{\frac{3 \tau  t^2}{2}+1}$, $Q=\frac{6 t^2 \tau ^2}{\left(\frac{3 \tau  t^2}{2}+1\right)^2}$. The behaviour of the scale factor from Fig. \ref{fig:model4} (a) confirms that the plot is vanishing at the bouncing region, which is shaded by light orange color. The Hubble parameter $H(t)$ from Fig. \ref{fig:model4}(b) shows Hubble parameter moves from negative region to positive region and is zero at the bouncing point. The EoS parameter lies in the phantom region at the bounce region and approaching towards $-1$ at early and late-time. This analysis implies that the $f(Q, L_m)$ gravity with the generalised power law model can succesfully address the matter bounce phenomenon. 
The cosmological viability of these models are tested using wel-known tool which energy conditions. The behaviour of all the energy conditions at each model have been investigated in the next section.

\section{Energy Conditions}\label{energy_conditions}

\begin{figure}[H]
\centering

\begin{subfigure}{0.45\textwidth}
\centering
\includegraphics[width=\linewidth]{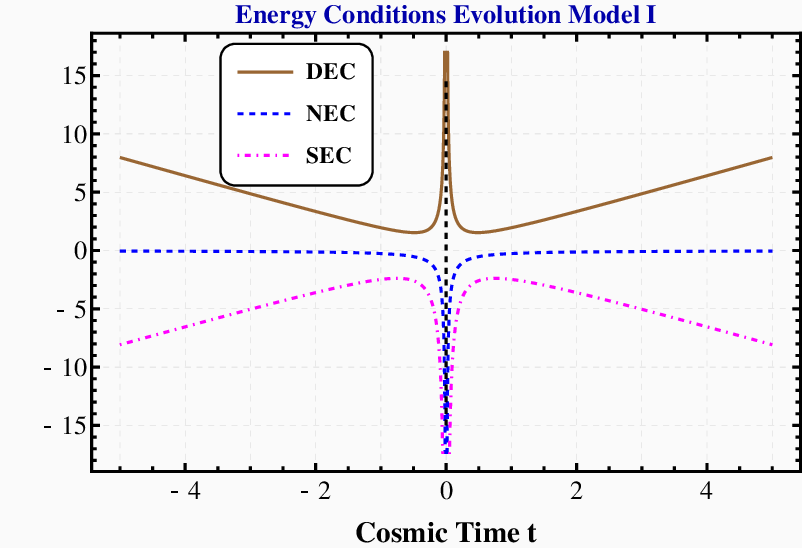}
\caption{Model I: Symmetric Bounce}
\end{subfigure}
\hfill
\begin{subfigure}{0.45\textwidth}
\centering
\includegraphics[width=\linewidth]{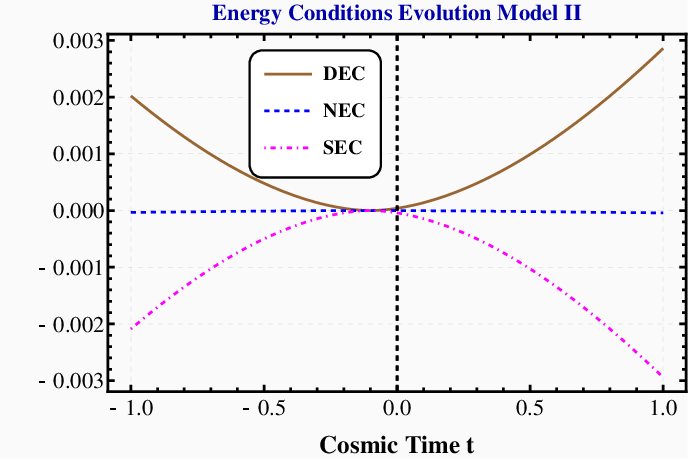}
\caption{Model II: Superbounce}
\end{subfigure}

\vspace{0.5cm}

\begin{subfigure}{0.45\textwidth}
\centering
\includegraphics[width=\linewidth]{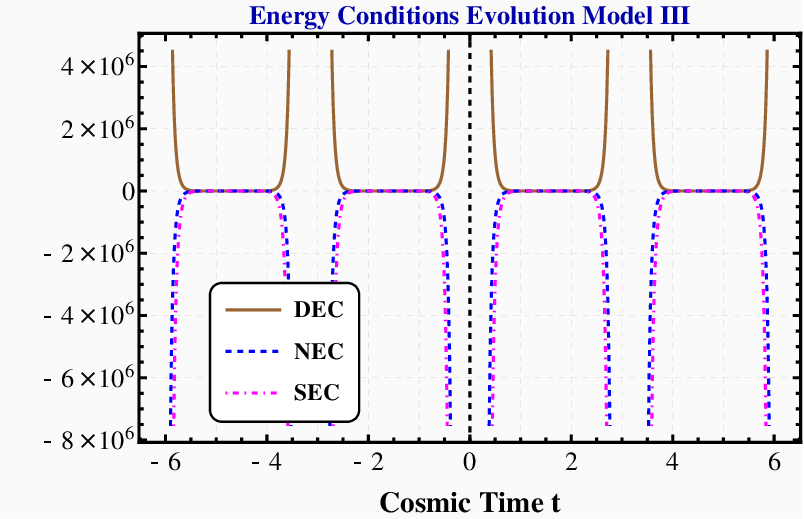}
\caption{Model III: Oscillatory Bounce}
\end{subfigure}
\hfill
\begin{subfigure}{0.45\textwidth}
\centering
\includegraphics[width=\linewidth]{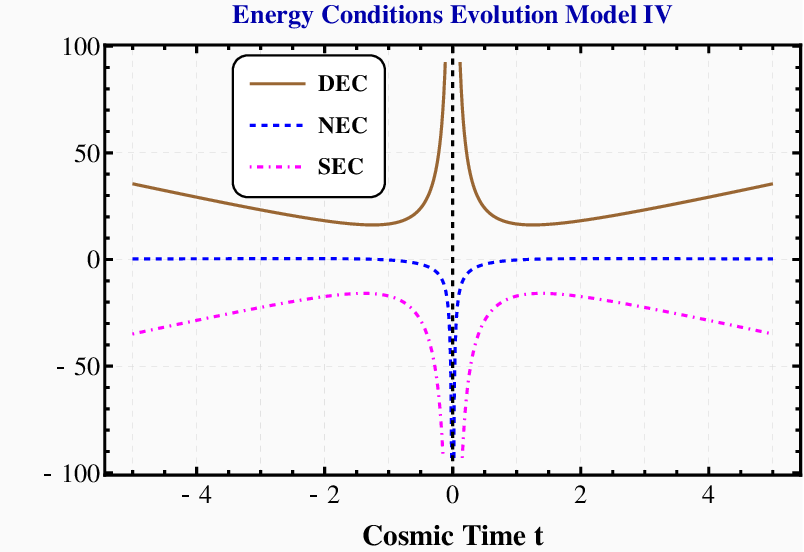}
\caption{Model IV: Matter Bounce}
\end{subfigure}

\caption{Evolution of the energy conditions for the four bouncing cosmological models: (a) symmetric bounce, (b) superbounce, (c) oscillatory bounce, and (d) matter bounce.}
\label{fig:energy_conditions_models}
\end{figure}

In this section, we analyze the behavior of the classical energy conditions corresponding to the bouncing cosmological models considered in this work. The evolution of the energy conditions for the four different models is illustrated in Fig.~\ref{fig:energy_conditions_models}. The general forms for energy conditions are \cite{CAPOZZIELLO201899}.
Energy conditions serve as a framework of constraints governing the relationship between energy density and pressure. These conditions ensure the physically consistent behavior of spacetime, dictate that energy density must be non-negative, and uphold the attractive nature of gravity. Specifically, they impose limitations on certain linear combinations of pressure and energy density, which cannot be negative. This plays a critical role in the investigations in late time cosmology \cite{PhysRevD.75.083523}, the study of wormholes \cite{Bejarano2017}, and the thermodynamics of black holes \cite{Singh2022}. The derivation of these conditions is rooted in Raychaudhuri’s equation \cite{Raychaudhri_55}. The general form of the four fundamental energy conditions is outlined as follows:
\begin{itemize}
    \setlength{\itemsep}{4pt}  % Adjust spacing between items
    \setlength{\parskip}{0pt}
    \setlength{\parsep}{0pt}

    \item Weak Energy Condition (WEC): $\rho_{\mathrm{eff}} \geq 0,\; p_{\mathrm{eff}} + \rho_{\mathrm{eff}} \geq 0$
    
    \item Null Energy Condition (NEC): $\rho_{\mathrm{eff}} + p_{\mathrm{eff}} \geq 0$
    
    \item Dominant Energy Condition (DEC): $\rho_{\mathrm{eff}} \geq \left| p_{\mathrm{eff}} \right|$
    
    \item Strong Energy Condition (SEC): $\rho_{\mathrm{eff}} + 3p_{\mathrm{eff}} \geq 0$
\end{itemize}

To achieve a successful bounce, it is required to violate the NEC, as depicted in Fig.~\ref{fig:energy_conditions_models}. This violation is critical because it allows the EoS parameter, $\omega$, to fall into the phantom region ($\omega < -1$) during the bounce, enabling a transition from contraction to expansion \cite{Cai2007}. Consequently, the NEC violation is fundamental for facilitating the bounce and avoiding singularities. The graph shows a pronounced NEC violation around the bounce epoch, supporting the notion of a non-standard expansion phase for the universe. Furthermore, the SEC is also violated in this framework, as shown in Fig.~\ref{fig:energy_conditions_models}, which directly stems from the NEC violation. This SEC violation plays a significant role in the model's evolution within the phantom phase, where $\omega < -1$, indicating that gravitational interactions are repulsive rather than attractive. Such a violation is consistent with the observed accelerated expansion of the universe surrounding the bounce. In addition, Fig.~\ref{fig:energy_conditions_models} reveals that the DEC is satisfied throughout the cosmic evolution. This satisfaction implies that the energy density remains non-negative, thereby preserving causality, consistent with the behavior of a perfect fluid-like matter distribution \cite{Odintsov2016}. 

\section{Summary and Conclusion}\label{conclsion}
The bouncing cosmology plays a vital role as an alternative to the standard model of cosmology. This phenomenon refers to addressing the cosmic expansion history and fate without dealing with the initial singularity in an alternative way to the $\Lambda$CDM model. In this study, we have explored modified formalism in nonmetricity gravity with the matter coupling $f(Q, L_m)=\beta +\alpha  L_m Q^{\mu }-\frac{Q}{2}$ \cite{Hazarika:2024alm, Myrzakulov_2025_EC}. The scope of unexplored cosmological bouncing solutions has been explored. The four well-known bouncing solutions, Model I symmetric bounce, Model II Superbounce, Model III Oscillatory, and Model IV matter bounce scenario, have been investigated. The investigation includes the behaviour of the scale factor, the Hubble parameter, and the EoS parameter. One can refer to Figs. \ref{fig:model1}-\ref{fig:model4}. The Hubble parameter appears from the negative to the positive region in each case, supporting bouncing behaviour. The energy conditions, which act as stability conditions, are analysed for each of the models discussed above. It has been observed from Fig. \ref{fig:energy_conditions_models}, at the bounce epoch, NEC violation supports the EoS parameter in the phantom region in all the cases investigated in this study. Moreover, there is a validation of the DEC and a violation of the SEC. These behaviours of the energy conditions perfectly reflect the model explaining the bouncing cosmology phenomenon in the modified non-metricity gravity $f(Q, L_m)$ formalism. 
\section*{Acknowledgements}
The authors acknowledge that they conducted this research without financial support from any public, commercial, or not-for-profit funding agency. 
\bibliographystyle{utphys}
\bibliography{biblio}

\end{document}